\documentclass{article}
\usepackage{kmn,graphics}

\newcommand{\eg}{{\sl e.g.}}

\newcommand{\etal}{{\sl et al. }}

\newcommand{\ergs}{$\,$erg$\,$s$^{-1}$}

\title[Spectroscopic Identification of the IR Counterpart to GX5-1]
{Spectroscopic Identification of the Infrared Counterpart to GX5-1}
\author[R.M. Bandyopadhyay et al.]{
R.M. Bandyopadhyay,$^{1}$\thanks{email:rmb@astro.ox.ac.uk} T. Shahbaz,$^{2}$ and P.A. Charles$^{3}$\\ 
$^{1}$Department of Astrophysics, University of Oxford, Keble Road, Oxford, OX1 3RH, UK \\
$^{2}$Instituto de Astrofisica de Canarias, C/Via Lactea, s/n, 38200 La Laguna, Tenerife, Spain\\
$^{3}$Department of Physics \& Astronomy, University of Southampton, Southampton SO17 1BJ, UK\\
}

\begin{document}

\maketitle

\begin{abstract}

\noindent
Using CGS4 on UKIRT, we have obtained a 1.95-2.45 $\mu$m infrared
spectrum of the primary candidate counterpart to the bright $Z$ LMXB
GX5-1.  IR photometry by Naylor, Charles, \& Longmore (1992) and the
astrometry of Jonker \etal (2000) had previously identified this star
as the most likely counterpart to GX5-1.  The spectrum presented here
clearly shows Brackett $\gamma$ and He lines in emission, for the
first time confirming the identity of the counterpart.  Similar to our
previous spectroscopy of the $Z$ source LMXBs Sco X-1 and Sco X-2
(Bandyopadhyay \etal 1999), the $K$-band spectrum of GX5-1 shows
emission lines only.  We briefly discuss the implications of this
spectrum for the nature of the $Z$ sources.

\end{abstract}


\section{Introduction}
In low-mass X-ray binaries (LMXBs), mass is transferred from a
late-type star to its compact companion, either a neutron star
or a black hole, via an accretion disc.  These systems can be
classified according to location within the Galaxy, accretion
characteristics, and luminosity (van Paradijs \& McClintock 1995).  A
group of persistent X-ray sources located within $15^{\circ}$
longitude and $2^{\circ}$ latitude of the Galactic Centre are known as
the ``galactic bulge sources'' (GBS; Warwick \etal 1988), and are
among the most luminous X-ray sources in the Galaxy (typical $L_{X}
\sim 10^{38}$ \ergs).  However, the GBS remain the most poorly
understood group of LMXBs, due to the heavy obscuration in the
direction of the Galactic bulge which makes optical study nearly
impossible.  The GBS are thought to be neutron star LMXBs, and
quasi-periodic oscillations (QPOs) have been detected in several
systems (\eg van der Klis 1989).  Attempts to detect orbital
variability in the GBS have generally been unsuccessful, suggesting
that their periods may be longer than those of canonical LMXBs
(Charles \& Naylor 1992).  On the basis of their X-ray colour-colour
diagrams, neutron star LMXBs have also been divided into two classes,
known as $Z$ and {\it atoll} sources (see van der Klis 1995 for a
review).  In this scheme, six of the brightest neutron star LMXBs
(which includes several of the GBS) are classified as $Z$ sources,
while the remainder fall into the atoll category.  Recently this
distinction has come into question, as several atoll sources
show $Z$-type colour-colour diagrams when examined on a sufficiently
long timescale (Muno \etal 2002, Gierlinski \& Done 2002).
Nevertheless clear differences remain between the X-ray luminosity and
spectral evolution of the canonical $Z$ sources and that of the atolls.

The infrared provides us with an ideal window for observing these
highly obscured systems.  Observing in the IR has two primary
advantages: the late-type secondaries in LMXBs are brighter relative
to the accretion discs, and, more importantly for the GBS, the ratio
of $V$- to $K$-band extinction is $\sim$9 ($A_{K}/A_{V} =$ 0.112;
Rieke \& Lebofsky 1985).  This advantage is particularly crucial in
the case of the $Z$ sources, as only two of the six (Sco X-1 and Cyg
X-2) have optically identified counterparts.  The low galactic
latitude of the remaining four $Z$ LMXBs ensures that they are so
heavily extincted in the optical that their counterparts can {\it
only} be observed in the IR.  We have undertaken a long-term program
of IR observations of X-ray binaries (XRBs), beginning with the
discovery via colours or variability of candidates for the
counterparts to the X-ray sources using precise X-ray and radio
locations (Naylor, Charles, \& Longmore 1991, hereafter NCL91).  Using
IR spectroscopy, we then attempt to definitively identify the
counterparts to the X-ray sources via detection of accretion
signatures in their spectra and also place constraints on the nature
of the mass donor stars.  We have obtained IR spectra of a number of
systems, including Sco X-1, GX13+1, and Sco X-2; our data suggests
that the secondaries in these LMXBs may be evolved rather than
main-sequence stars (Bandyopadhyay \etal 1997 and 1999, hereafter
Papers I and II respectively).

After Sco X-1, at X-ray wavelengths GX5-1 is the brightest of the
persistent LMXBs.  It exhibits a variety of QPOs and, like all $Z$
LMXBs, is a radio source (Fender \& Hendry 2000).  Our previous
spectroscopy of two stars within the {\it Einstein} X-ray error circle
(stars 502 and 503, as designated by NCL91; see Figure 1) showed no
evidence for the signature emission lines we would expect in an LMXB
(Paper II).  Astrometry by Jonker \etal (2000) produced an improved IR
position for star 513 which was nearly coincident with the ATCA radio
position of GX5-1 (Berendsen \etal 2000).  Therefore of the candidate
stars identified by NCL91, star 513 remained as the prime candidate
for the IR counterpart to GX5-1. In this Letter we present $JHK$
photometry and a $K$-band spectrum of star 513, confirming that this
star is the IR counterpart to GX5-1.

\section{Observations and Data Analysis} 

\subsection{Photometry}

We obtained $JHK$ images of the GX5-1 field using the UKIRT Fast-Track
Imager (UFTI) on the 3.8-m United Kingdom Infrared Telescope on Mauna
Kea on the nights of 2001 July 9 and 10 UT.  UFTI uses a 1024x1024
HAWAII HgCdTe array and has a plate scale of 0.091 arcsec
pixel$^{-1}$, giving a field of view of 93.06 arcsec$^{2}$; see Roche
\etal (2002) for further information about this instrument.  On both
nights, a series of five consecutive 60-second images were taken in
each filter in a jitter pattern so that the group could be mosaiced
and used to create a flat field.

The initial reduction of the images was performed using online data
reduction (ORAC\_DR) scripts (Wright \& Leggett 1997).  This process
includes dark subtraction, flat field creation and division, masking
of bad pixels, and mosaicing of the jittered frames into the final
image.

The data were analyzed using standard IRAF aperture photometry
routines and an 8-pixel aperture, chosen to maximize the target signal
relative to the sky background.  Relative photometry was performed
using several field standards.  Finally, absolute photometry of the
local standards was performed.  A journal of the IR observations with
the measured $JHK$ magnitudes of GX5-1 is presented in Table 1.  The
$H$ and $K$ magnitudes that we find for star 513 are consistent with
those observed by Jonker \etal (2000), and there is little evidence
for significant variability between the two nights.

\begin{table}
\caption{$JHK$ magnitudes of GX5-1 (star 513)}
\begin{center}
\begin{tabular}{lcc}\hline
UT Date   &  Filter  & Magnitude     \\\hline 
9 July    &  $J$     & 14.68$\pm$0.06    \\ 
          &  $H$     & 14.11$\pm$0.07   \\
          &  $K$     & 13.56$\pm$0.05   \\ 
10 July   &  $J$     & 14.79$\pm$0.05   \\ 
          &  $H$     & 14.07$\pm$0.06  \\
          &  $K$     & 13.53$\pm$0.05   \\ \hline
\end{tabular}
\end{center}
\end{table}

\subsection{Spectroscopy}

On 2001 July 10 UT we obtained $K$-band (1.95--2.45 $\mu$m) spectra of
star 513 using the Cooled Grating Spectrometer (CGS4) on UKIRT, with
an on-source integration time of 3840 s.  The slit width was 0.61
arcseconds which corresponds to 1 pixel on the detector.  The 40 l/mm
grating was used with the 300mm camera and the 256$\times$256 pixel
InSb array, yielding a spectral dispersion of 2.5 nm pixel$^{-1}$ ($R
\sim 880$ at 2.2$\mu$m).  Target observations were bracketed by
observations of A-type stars for removal of telluric atmospheric
features.

The standard procedure of oversampling was used to minimise the
effects of bad pixels (Wright 1995).  The spectra were sampled over
two pixels by mechanically shifting the array in 0.5 pixel steps in
the dispersion direction.  We employed the non-destructive readout
mode of the detector in order to reduce the readout noise.  In order
to compensate for the fluctuating atmospheric emission lines we took
relatively short exposures and nodded the telescope so that the object
spectrum switched between two different spatial positions on the
detector.  Details of the design and use of CGS4 can be found in
Mountain \etal (1990).

The CGS4 data reduction system performs the initial reduction of the
2D images.  These steps include the application of the bad pixel mask,
bias and dark subtraction, flat field division, interlacing
integrations taken at different detector positions, and co-adding and
subtracting the nodded images (see Daly \& Beard 1994).  Extraction of
the 1D spectra, wavelength calibration, and removal of the telluric
atmospheric features was then performed using IRAF.  A more detailed
description of the data reduction procedure is provided in Paper I.

\section{Discussion}

Our $JHK$ images of the GX5-1 field are shown in Figure 1; the stars
are labelled using the designations of NCL91.  Our $K$-band spectrum
of star 513 is shown in Figure 2; strong emission lines of HI and HeI
are the dominant features, directly confirming the identification of
star 513 as the IR counterpart to GX5-1.  However, no absorption lines
from the secondary can be distinguished.  A list of identified lines
is given in Table 2.

\begin{table}
\caption{Emission lines detected in GX5-1}
\begin{center} 
\begin{tabular}{lcc} \hline
Line          	&  $\lambda$   & Equivalent width 	\\
	      	   & ($\mu$m) $^{a}$ & (\AA) 	\\ \hline
Br $\delta$   	 &   1.944     & -29.9$\pm$0.5 \\ 
HeI	        	 &   2.057     & -14.1$\pm$0.6 \\
Br $\gamma$   	 &   2.165     & -45.3$\pm$0.5 \\ \hline
\end{tabular} 
\end{center}
\noindent $^{a}$ Typical wavelength calibration errors are
approximately $\pm$0.001 $\mu$m. \\
\end{table}

Of the six known $Z$ sources, GX5-1 is the fourth for which an IR
counterpart has been spectroscopically identified.  Of the remaining
two $Z$ LMXBs, GX17+2 has had a probable counterpart identified on the
basis of HST and Keck IR photometry (Deutsch \etal 1999, Callanan
\etal 1999), while to date no candidate counterpart has been
identified for GX340+0.  In Papers I and II we presented $K$-band
spectroscopy of the $Z$ sources Sco X-1 and Sco X-2, which, as in the
spectrum of GX5-1 presented here, show only emission lines from the
accretion disc and/or heated face of the secondary star.  Modelling of
the Sco X-1 spectrum, including the accretion disc contribution,
suggested that the mass donor is likely to be an evolved star of
spectral type earlier than G5.  Similarly, the lack of optical or IR
spectroscopic features associated with the secondary in Sco X-2
indicates that the mass-donating star is also of an earlier type than
the K/M secondaries expected in atoll LMXBs (Paper II).  Although the
orbital period of Sco X-2 has not been conclusively determined, all
the candidate periodicities proposed to date are consistent with an
evolved rather than main sequence companion.  Finally, the secondary
in Cyg X-2 has been definitively identified as an evolved A9 star
(Casares, Charles \& Kuulkers 1998).  These results led us to
hypothesize that the secondaries in the six $Z$ sources may all be
{\it early-type evolved stars}.  The spectrum of GX5-1 presented here
shows no absorption features from the secondary which would contradict
this hypothesis.  Thus the spectra of all four of the six $Z$ sources
with spectroscopically identified counterparts are consistent with
mass donors of type G or earlier, rather than ``canonical'' K/M
secondaries.

However, the $Z$ sources are among the brightest persistent XRBs in
the Galaxy, and as such we may expect their accretion discs to be
exceptionally bright and possibly exceptionally large (\eg if their
orbital periods prove to be substantially longer than those of
canonical LMXBs).  Therefore even in the IR it is possible that
absorption features from the secondary are obscured by the luminous
accretion disc.  Although our previous modelling of the Sco X-1
spectrum indicated that even with a large disc contribution in the
$K$-band we would expect to see absorption lines if the companion was
a K/M star (Paper II), this issue can only be resolved with very high
S/N ($>$100) spectroscopy.  For example, in the case of GRS~1915+105,
which is also an extremely bright quasi-persistent X-ray source,
$K$-band spectra with moderate S/N ($\sim$30) also showed no
absorption features (Mirabel \etal 1997, Marti \etal 2000).  However,
spectra of GRS~1915+105 obtained on the VLT with long exposure times
and thus much higher S/N revealed the CO band absorption features of
the secondary (Greiner \etal 2001).  Therefore the definitive test of
our hypothesis on the nature of the $Z$ source mass donors will be to
obtain similarly high S/N spectra of Sco X-1, Sco X-2, and GX5-1.

\section{Acknowledgements}

A portion of this work was performed while RMB was supported by a
National Research Council Research Associateship at the Naval Research
Laboratory.

\newpage
\clearpage

\onecolumn
\begin{figure} 
\begin{center}
\includegraphics{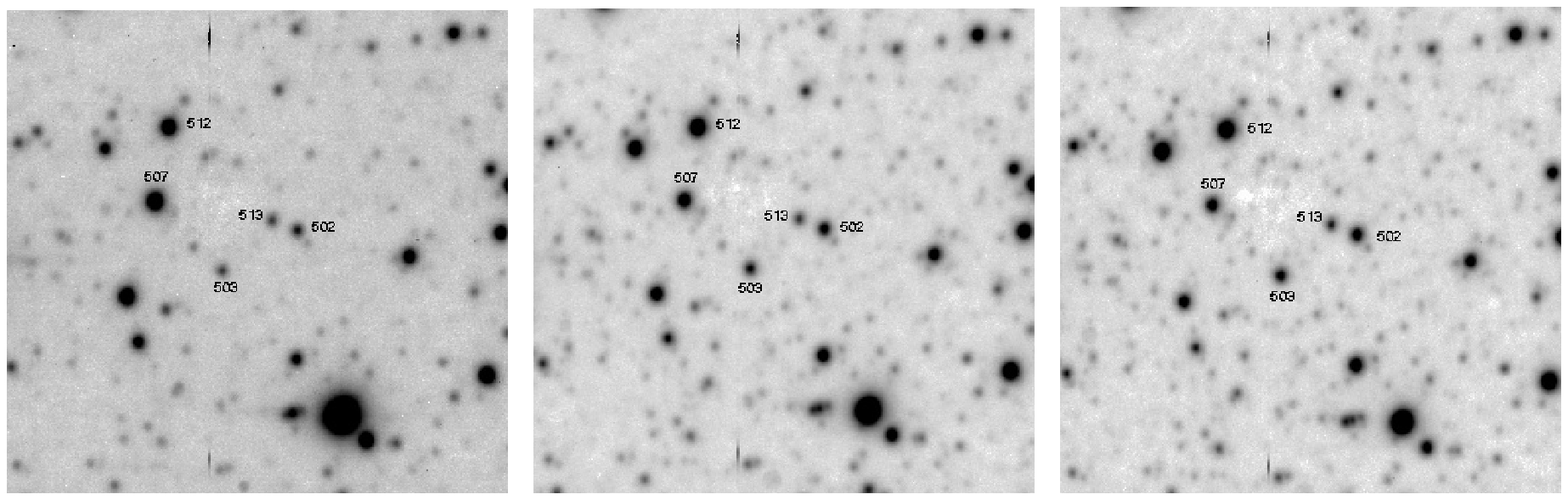}
\caption{$J$, $H$, and $K$ -band images (left to right) of the GX5-1 field obtained on 10 July 2001 with UFTI on UKIRT.  Stars are labelled using the number designations of NCL91.  The magnitude of star 513 is $K$ = 13.53$\pm$0.05.}
\end{center}
\end{figure} 

\begin{figure} 
\begin{center}
\scalebox{0.5}{\includegraphics{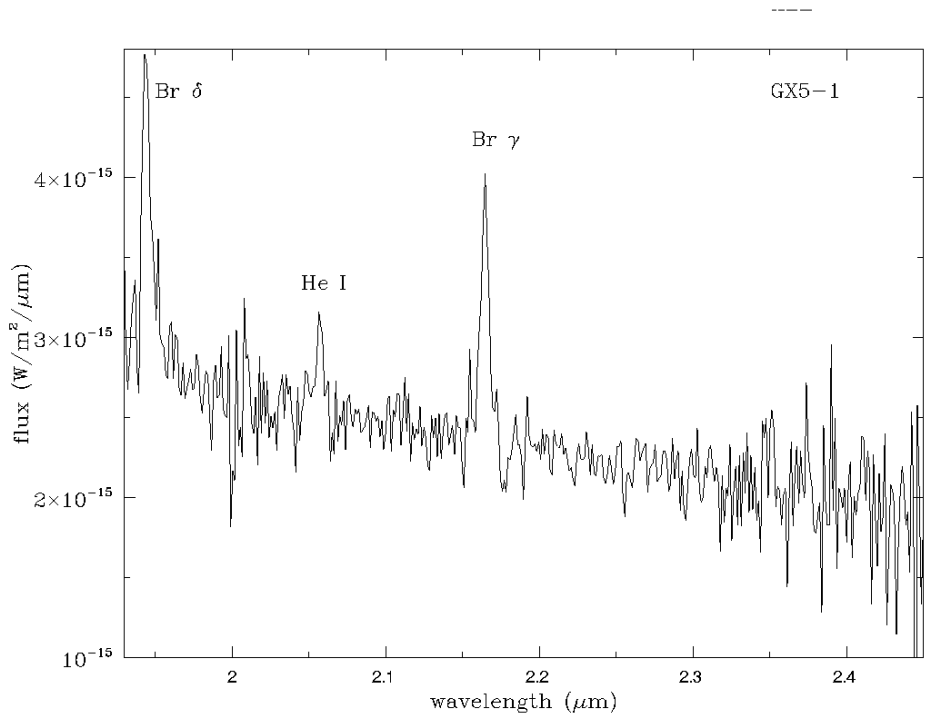}}
\caption{$K$-band spectrum of star 513, obtained on 10 July 2001 with CGS4 on UKIRT.  Strong H and HeI emission lines are clearly visible, confirming the identification of this star as the IR counterpart to GX5-1.}
\end{center}
\end{figure} 

\end{document}